\documentclass[aps,prl,reprint, superscriptaddress]{revtex4-2}
\usepackage{bbold}
\usepackage{comment}
\usepackage[colorlinks, citecolor=red]{hyperref}
\usepackage{hyperref}
\usepackage[utf8]{inputenc}
\usepackage[T1]{fontenc}    %
\usepackage[english]{babel}
\usepackage[pdftex]{graphicx}
\usepackage{amsmath,amssymb,amsthm,bbm, mathtools} %
\usepackage{mathrsfs}
\usepackage{xcolor}
\usepackage[normalem]{ulem}
\usepackage{multirow}

\begin{document}
\title{Electrically Tuneable Variability in Germanium Hole Spin Qubits}
\author{Edmondo Valvo}
\altaffiliation{These authors contributed equally to this work.}
\email{e.valvo@tudelft.nl}
\affiliation{QuTech and Kavli Institute of Nanoscience, Delft University of Technology, Delft, The Netherlands}
\author{Michèle Jakob}
\altaffiliation{These authors contributed equally to this work.}
\email{m.jakob@tudelft.nl}
\affiliation{QuTech and Kavli Institute of Nanoscience, Delft University of Technology, Delft, The Netherlands}
\author{Patrick Del Vecchio}
\affiliation{QuTech and Kavli Institute of Nanoscience, Delft University of Technology, Delft, The Netherlands}
\email{@tudelft.nl}
\author{Maximilian Rimbach-Russ}
\affiliation{QuTech and Kavli Institute of Nanoscience, Delft University of Technology, Delft, The Netherlands}
\email{m.f.russ@tudelft.nl}
\author{Stefano Bosco}
\affiliation{QuTech and Kavli Institute of Nanoscience, Delft University of Technology, Delft, The Netherlands}
\email{s.bosco@tudelft.nl}
\date{\today}

\footnotetext{These authors contributed equally to this work}

\begin{abstract}
Hole spin qubits in planar germanium heterostructures are frontrunners for scalable semiconductor quantum computing. However, their current performance is mostly limited by large dot-to-dot variability that leads to uncontrolled qubit energies and random tilts in the spin quantization axis. Here, we propose a systematic and local method to engineer the spin qubit response by imprinting a controlled anisotropy in the quantum dot confinement, enabling on-demand electric g-tensor control. In particular, we find that both the quantum-dot size and asymmetry allow electrical tuning of the g-tensor and significantly suppress magnitude and angular variability of the spin response for selected magnetic field directions. We confirm this behavior by analyzing single-disorder realizations and statistical ensembles in state-of-the-art strained and unstrained germanium channels, showing that the latter provides an optimal path for $g$-tensor engineering. Our results provide practical design principles for on-demand control of the spin response and mitigating variability, paving the way towards large-scale germanium-based quantum computers.
\end{abstract}
\maketitle

\paragraph{Introduction.--} 
Spin qubits confined in quantum dots (QDs)~\cite{lossQuantumComputationQuantum1998} are leading candidates for scalable semiconductor quantum computing~\cite{burkardSemiconductorSpinQubits2023,stanoReviewPerformanceMetrics2022,vandersypenInterfacingSpinQubits2017,philipsUniversalControlSixqubit2022,xueQuantumLogicSpin2022,millsTwoqubitSiliconQuantum2022,zwerverShuttlingElectronSpin2023,desmetHighfidelitySinglespinShuttling2025,takedaQuantumErrorCorrection2022,noiriShuttlingbasedTwoqubitLogic2022,wuSimultaneousHighFidelitySingleQubit2025,steinackerIndustrycompatibleSiliconSpinqubit2025,gonzalez-zalbaScalingSiliconbasedQuantum2021,petitUniversalQuantumLogic2020,siegelSnakesPlaneMobile2025,PhysRevA.109.032433,svastitsReadoutSweetSpots2025a,liFlexible300Mm2020,kloeffelProspectsSpinBasedQuantum2013,fangRecentAdvancesHolespin2023,maurandCMOSSiliconSpin2016,Hendrickx2024,tidjaniThreeDimensionalArrayQuantum2025}. Among them, hole spins in planar germanium (Ge) heterostructures stand out for their relaxed fabrication requirements, micromagnet-free operations, and high-fidelity all-electrical single- and two-qubit gates~\cite{zhangUniversalControlFour2025,vanriggelenPhaseFlipCode2022,Wang2024Hopping,hendrickxFourqubitGermaniumQuantum2021,jirovecSinglettripletHoleSpin2021,wangUltrafastCoherentControl2022,liuUltrafastElectricallyTunable2023,lawrieSimultaneousSinglequbitDriving2023,Scappucci2020,Stehouwer2025,vorreiterPrecisionHighspeedQuantum2025,jirovecManybodyInterferometrySemiconductor2025,saez-mollejoExchangeAnisotropiesMicrowavedriven2025,johnTwodimensional10qubitArray2025a,jirovecMitigationExchangeCrosstalk2025,ivlevOperatingSemiconductorQubits2025,farinaSiteresolvedMagnonTriplon2025}.
Their compatibility with superconductors and strong spin-orbit interaction (SOI), especially in unstrained channels, further make Ge a promising platform for hybrid superconducting-semiconducting quantum devices~\cite{costaBuriedUnstrainedGermanium2025,boscoSqueezedHoleSpin2021,mauroHoleSpinQubits2025}.

A key challenge, however, is the high susceptibility of their spin response to the local electrostatic and strain configuration, which in current devices tilts the quantization axes of neighboring QDs in highly unpredictable directions~\cite{martinez2025variabilityholespinqubits,Martinez2022,stehouwerGermaniumWafersStrained2023,sangwanImpactSurfaceTreatments2025,paqueletwuetzReducingChargeNoise2023,elsayedLowChargeNoise2024}. Although such rotations permit baseband control and alternative two-gate schemes~\cite{boscoExchangeOnlySpinOrbitQubits2024a,rimbach-russGaplessSingleSpinQubit2025,seidlerSpatialUniformityGtensor2025,jirovecDynamicsHoleSingletTriplet2022}, their unpredictability so far prevents systematic scaling. Controlling both the magnitude and angular variability of the spin response is therefore essential for scalable quantum processors.
Stabilizing the $g$-tensor amplitude enables systematic sweet spots ~\cite{ bassiOptimalOperationHole2024,carballidoCompromisefreeScalingQubit2025, PRXQuantum.2.010348,Wang2024Modelling, Terrazos2021} 
that reduce qubit dephasing, while suppressing angular variability is crucial for reliable qubit operations~\cite{Wang2024Hopping,rimbach-russGaplessSingleSpinQubit2025,kp8s-py9m, geyerAnisotropicExchangeInteraction2024, PhysRevB.109.085303} 
and for long-range interconnects based on spin shuttling~\cite{vanriggelen-doelmanCoherentSpinQubit2024, ademiDistributingEntanglementDistant2025,matsumotoTwoqubitLogicTeleportation2025,seidlerConveyormodeSingleelectronShuttling2022,xueSiSiGeQuBus2024} and spin-photon coupling~\cite{depalmaStrongHolephotonCoupling2024,janikStrongChargephotonCoupling2025,PhysRevLett.129.066801,yuStrongCouplingPhoton2023,peterssonCircuitQuantumElectrodynamics2012a,miCoherentSpinPhoton2018,samkharadzeStrongSpinphotonCoupling2018,landigVirtualphotonmediatedSpinqubittransmonCoupling2019,dijkemaCavitymediatedISWAPOscillations2025}.

\begin{figure}[!t]
    \centering
    \includegraphics[width=\linewidth]{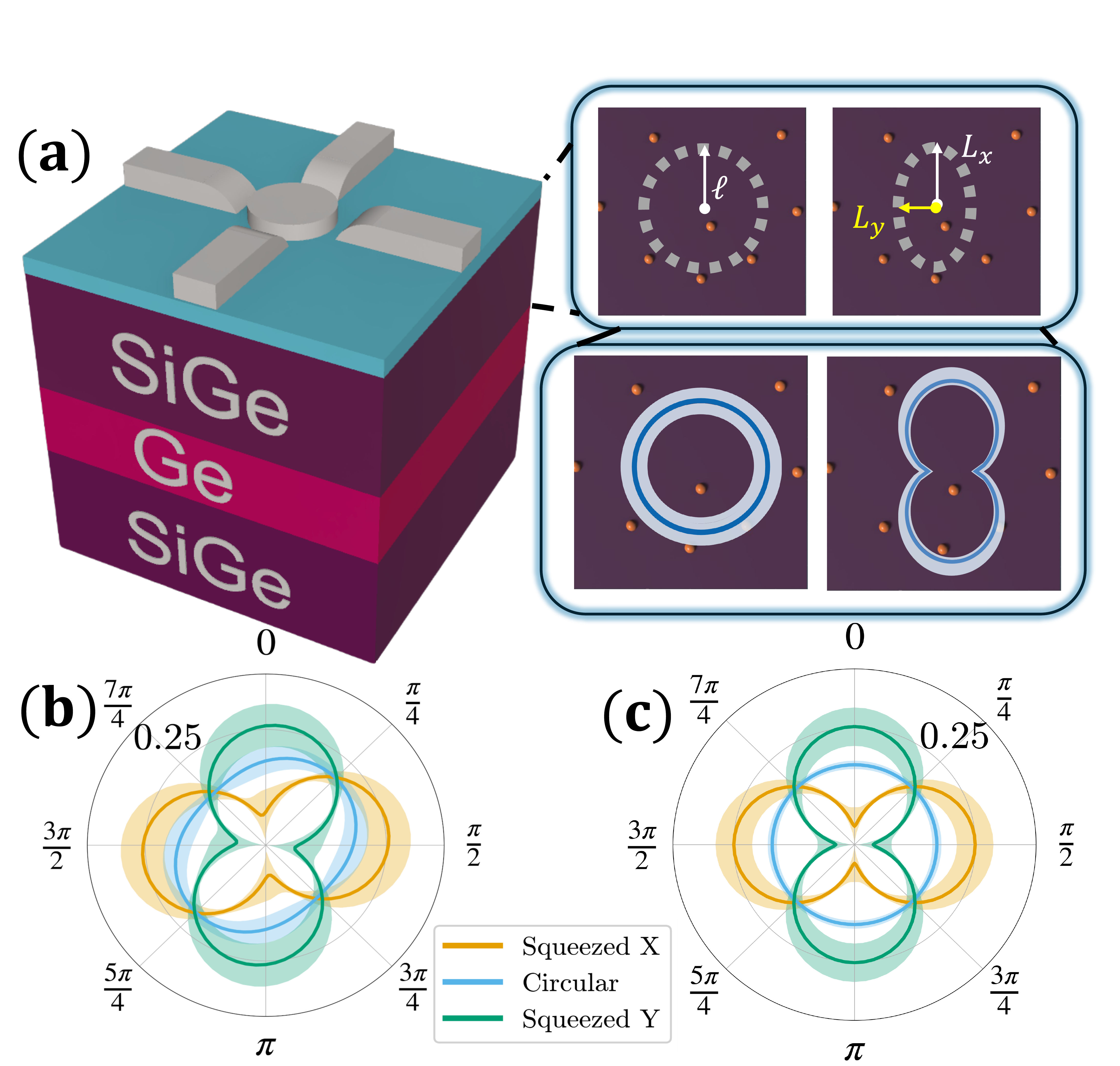}
    \caption{\textbf{Robust g-tensor engineering by squeezing.} (a) Sketch of planar Ge heterostructures with circular and squeezed QDs, including interface disorder (orange dots). Circular QDs yield isotropic spin response, with symmetry randomly broken by electrostatic disorder and fluctuations in the harmonic confinement. In contrast, squeezing QDs pins the $g$-tensor along the long axis, making it more robust against disorder.
    (b)-(c) Simulated $g$-tensors and standard deviations~\cite{footnoteFig1}, for strained (b) and unstrained (c) Ge, comparing circular and squeezed QDs oriented in different directions. In (b), uncontrolled long-range strain fluctuations pin the circular QD $g$-tensors and are partly compensated by squeezing. In (c), these fields are absent enabling reliable $g$-tensor engineering. 
    We used $\ell=40$~nm, $L_z=20$~nm, and long (short) in-plane lengths of $40$~nm ($20$~nm).     }
    \label{fig:placeholder}
\end{figure}  

In this work, we propose a scalable design strategy for Ge QDs that enables on-demand control over both the orientation and variability of the spin quantization axis. By electrically imprinting an anisotropy in the QD confinement, achieved by in-plane squeezing the QD potential away from circular symmetry, as sketched in Fig.~\ref{fig:placeholder}(a), we demonstrate a controlled alignment of the spin quantization axis even in the presence of inhomogeneous strain and electrostatic disorder. Our approach permits a scalable way to on-demand engineer g-tensors while suppressing their variability for selected in-plane magnetic-field directions.
By also analyzing the effects of QD size, we predict reduced variability for smoother confinements, consistent with circular QDs, and establish practical guidelines for achieving homogeneous and controllable spin behavior across large-scale hole-qubit arrays.

\begin{figure*}[htbp]
    \centering
    \includegraphics[width=\linewidth]{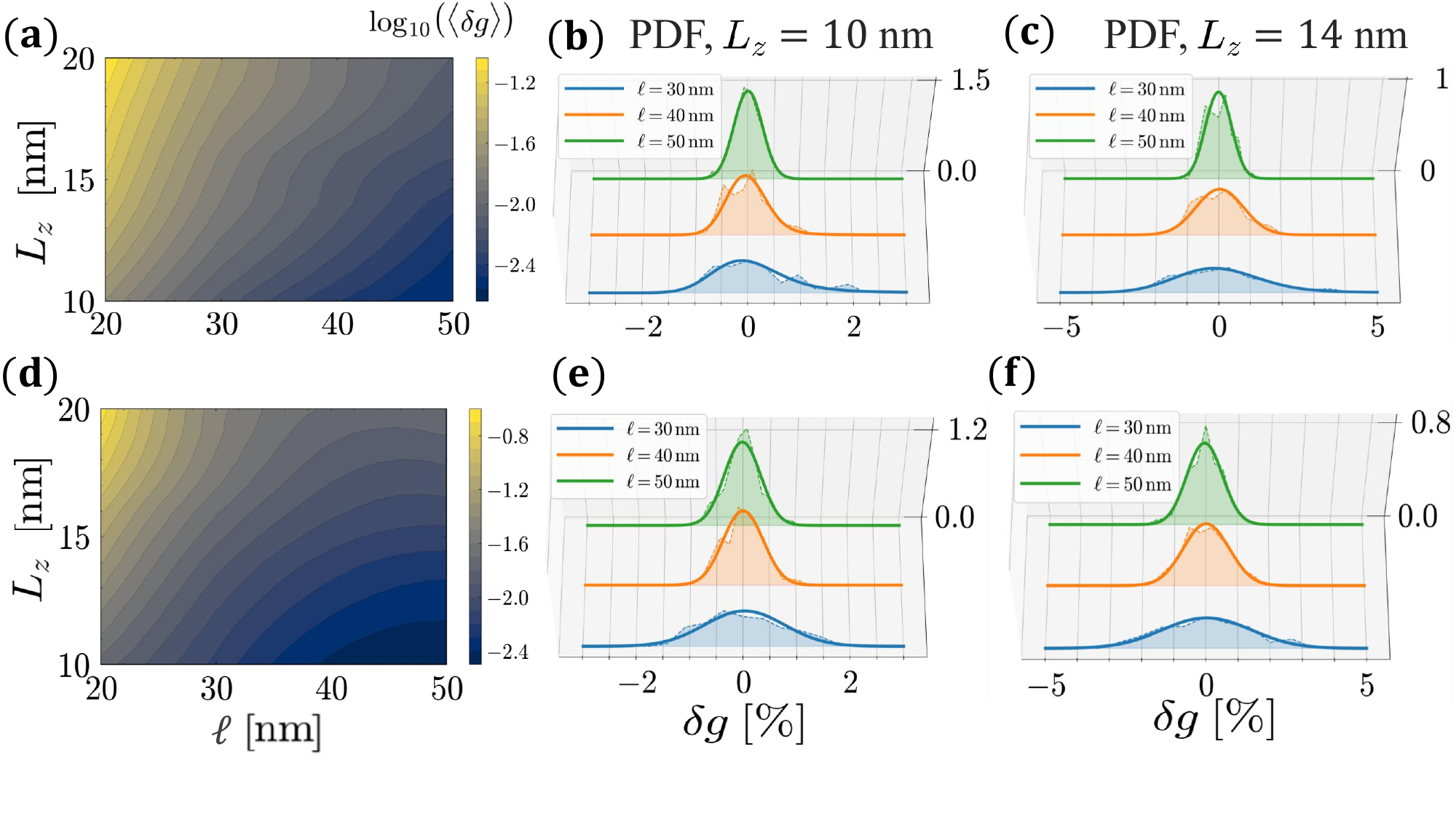}
    \caption{\textbf{Variability of circular QDs.}  (a)-(b)-(c) Standard deviation (a) and probability distribution functions of the relative $g$-factors $\delta g$ of circular QDs in strained Ge with $L_z=10$~nm (b) and   $L_z=14$~nm (c). We compare these quantities to unstrained Ge in (d)-(e)-(f). The results in both cases are obtained by averaging over 200 disorder configurations. For circular QDs, the variability is generally lower at smoother confinements both in-plane and out-of-plane, although the trend especially in unstrained Ge is non-monotonous.}
    \label{fig:g vs Lz}
\end{figure*}

\paragraph{Ge hole spin qubits.--} 
Without magnetic fields, holes in Ge are modeled by the Hamiltonian
\begin{equation}
\label{eq.H0}
H_0=H_{\text{LK}}+H_{\text{BP}}+V+V_D  \ .    
\end{equation}
Here, $H_{\text{LK}}$ and $H_\text{BP}$ denote the 4-band Luttinger–Kohn and Bir–Pikus Hamiltonians~\cite{Scappucci2020}, which describe the mixing of heavy and light holes at the $\Gamma$ point induced by the crystal momentum $\textbf{k}=-i\nabla$ and by the strain tensor $\varepsilon_{ij}$, respectively.
We include the homogeneous strain $\varepsilon^{(0)}$ arising from the mismatch of lattice constants between the Ge channel and the SiGe barriers and the inhomogeneous strain $\varepsilon^{(1)}$ imprinted by the metallic gates~\cite{lilesElectricalControlTensor2021,PhysRevB.104.235303,abadillo-urielHoleSpinDrivingStrainInduced2023,corley-wiciakNanoscaleMapping3D2023}. 
Explicitly, we consider
\begin{equation}
    \varepsilon_{ij}(x,y)= \varepsilon_{ii}^{(0)}\delta_{ij}+\sum_{m,n=-4}^4 \varepsilon^{(1)}_{mn} e^{2 \pi i\left({m x + n y} \right)/{L_p}} \ ,
\end{equation}
where we decomposed the local strain into a fourth-order Fourier series, which reflects the symmetry of the gates in Fig.~\ref{fig:placeholder}(a); $L_p$ is the radius of the plunger gate. 
We analyze two types of heterostructures: the typical strained Ge channel~\cite{corley-wiciakNanoscaleMapping3D2023,Stehouwer2025} with $\varepsilon_{xx}^{(0)}=\varepsilon_{yy}^{(0)}=-0.61 \%$ and  $\varepsilon_{zz}^{(0)}= 0.1 \%$~\cite{boscoSqueezedHoleSpin2021,sarkarEffectDisorderStrain2025} and the unstrained Ge channel with $\varepsilon_{ii}^{(0)}=0$, having enhanced SOI and reduced disorder~\cite{costaBuriedUnstrainedGermanium2025, boscoSqueezedHoleSpin2021, mauroHoleSpinQubits2025}.
More detail, including figures of the inhomogeneous strain patterns analyzed and explicit values of $\varepsilon_{ij}^{(1)}$, are provided in the Supplemental Material (SM)~\cite{supplemental_material}. 

The QD is defined by the potential $V =V_{z}-eF_z z +m\left( \omega_x^2 x^2 + { \omega_y^2} y^2\right)/2 $.
In the $z$-direction, the hole is confined by a hard-wall potential $V_{z}$ of width $L_z$. The gates define an electrostatic potential comprising an electric field $F_z$ in the $z$-direction and a harmonic in-plane potential with frequencies $\omega_i$ which are related to the respective confinement lengths by $L_i = \sqrt{{\hbar}/{m \omega_i}}$ with $i=x,y$. We use $m=m_0/\sqrt{\gamma_1^2 -\gamma_2^2}$, with $m_0$ being the electron mass and $\gamma_i$ being the bulk Ge Luttinger parameters~\cite{supplemental_material}.

To analyze the variability of the spin response, we also include the disorder potential $V_D= V_C+V_E$.  The first term 
\begin{equation}
\label{Coulomb}
    V_C=\frac{e^2}{4 \pi \epsilon_r}\sum_i \left(\frac{1}{|\textbf{r}-\textbf{r}_i|}-  \frac{1}{|\textbf{r}-(\textbf{r}_i+2z_m\textbf{e}_z)|} \right) 
\end{equation}
models the Coulomb potential of an ensemble of charge defects at the position $\textbf{r}_i$ fixed at the SiGe-Si oxide interface, assumed to be 50~nm away from the Ge channel, and screened by a top gate $z_m=7$~nm apart~\cite{shehataModelingSemiconductorSpin2023a}. Here, $\epsilon_r$ is the Ge dielectric constant. We consider the typical defect density of $10^{10}$~cm$^{-2}$~\cite{paqueletwuetzReducingChargeNoise2023}, which corresponds to $\sim 10$ randomly distributed charges in an area of $300\times 300$~nm$^2$. We focus  on negative charges at the interface, consistent with fixed charges at the oxide~\cite{berghuisSurfacePassivationGermanium2021}, but emphasize that, as explicitly shown in the SM~\cite{supplemental_material}, our results are valid also for positive charges. 
As a further source of variability, we introduce at each disorder iteration a stochastic electrostatic potential 
\begin{equation}\label{Shift and WF oscillation}
    V_E=- e\pmb{\delta}\textbf{F}\cdot \textbf{r} +  \frac{m\pmb{\delta\omega}^2\cdot \textbf{r}^2 }{2} \ ,
\end{equation}
that models possible different electrostatic tunings of different QDs. This potential is characterized by a random in-plane electric field $\pmb{\delta}\textbf{F}$  that shifts the wavefunction by up to $\pm 5$~nm from the center of the system and by a random modulation of the QD size  $\pmb{\delta\omega}$ that changes the size of the QD $L_i$ by $\pm10\%$. This modulation causes shifts in the electrochemical potential between $20$~$\mu$eV and $70$~$\mu$eV depending on the planar confinement strength.

The effect of a small applied magnetic field $\textbf{B}$ is captured by $ H_{\text{B}}= H_z-H_o$. This includes the Zeeman  $H_z= 2 \mu_B \textbf{B}\cdot(\kappa \textbf{J}  + q \textbf{J}^3)$ and orbital contribution $H_\text{o}=2\mu_B [\gamma_3\{\textbf{A}\cdot\textbf{J},\textbf{k}\cdot \textbf{J}\}+(\gamma_2-\gamma_3)\{\textbf{A},\textbf{k}\}\cdot\textbf{J}^2]$~\cite{PhysRevB.97.235422,PhysRevB.105.075308,Terrazos2021} arising from the dynamical momentum  $\pmb{\pi}=\hbar\textbf{k} +e \textbf{A}$ with vector potential $\textbf{A}= (B_x y-B_y x) \textbf{e}_z$ and with anticommutator $\{A,B\}=AB+BA$. We restrict our analysis to in-plane magnetic fields, the typical experimental choice to suppress hyperfine interactions~\cite{PhysRevB.78.155329,PhysRevLett.127.190501}.

To extract the properties of the qubit, we follow the standard approach~\cite{Venitucci2018,Wang2024Modelling,PRXQuantum.2.010348,PhysRevLett.129.066801,sharmaGfactorSymmetryTopology2024,AbragamBleaneyParamagenticResonance} and we construct the $g$-tensor  of the hole spin by projecting $H_\text{B}$ onto the numerically-computed two-fold degenerate groundstate of $H_0$ in Eq.~\eqref{eq.H0}.
We analyze the variability of the qubit by comparing the $g$-tensor obtained for different realizations of the electrostatic disorder $V_D$ produced by different random ensembles of charge defects. We also quantify the effect of electrostatic disorder by introducing the relative variations 
\begin{equation}
\label{eq:dg}
    \delta g = \frac{\lVert (g_0-g_N)\cdot \textbf{B} \rVert }{\lVert g_0\cdot \textbf{B}\rVert} \ ,  \ \ \delta\phi=\phi_N^{\text{Max}} - \phi_0^{\text{Max}}
\end{equation}
of the amplitude and tilt of the $g$ tensor with and without electrostatic disorder, $g_N$ and $g_0$, respectively. The relative amplitude variation $\delta g$ directly determines the variation of the qubit frequency, while the angular tilt $\delta\phi$ is critical for qubit operations~\cite{Wang2024Hopping,geyerAnisotropicExchangeInteraction2024,saez-mollejoExchangeAnisotropiesMicrowavedriven2025} and shuttling~\cite{vanriggelen-doelmanCoherentSpinQubit2024}. We compute $\delta\phi$ as the difference of the in-plane magnetic field angle $\phi_B$ corresponding to the maxima of the in-plane g-tensors $g_N$ and $g_0$. Typical $g_0$ amplitudes and corresponding Zeeman energy variations are reported in the SM~\cite{supplemental_material}.
 
\begin{figure}[htbp]
    \centering
    \includegraphics[width=\linewidth]{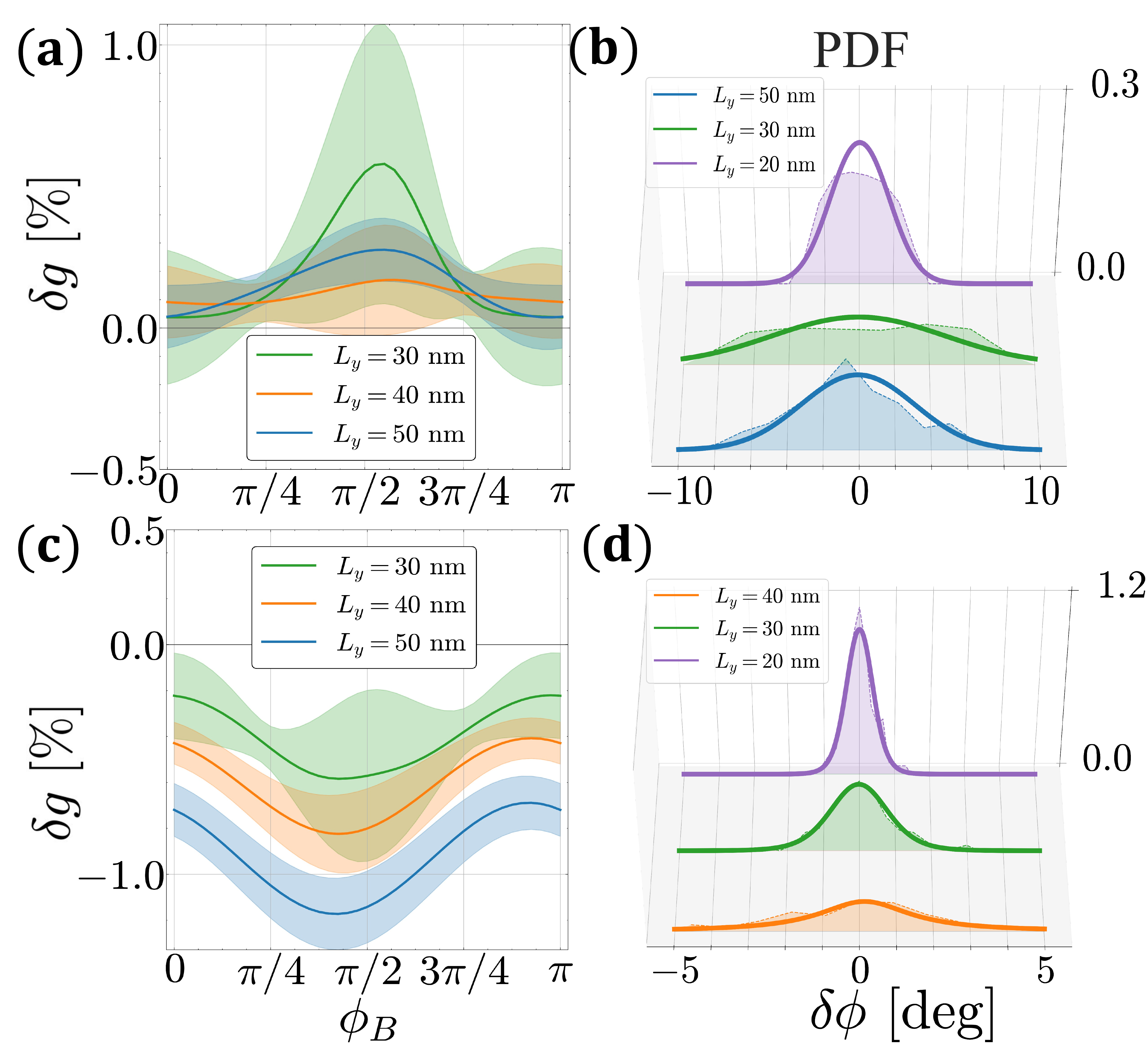}
    \caption{\textbf{Variability of squeezed QDs.}   (a) Simulated variability of the relative $g$-factor $\delta g$ for different $\textbf{B}$ field directions $\phi_B$ and (b) probability density function of the tilt angle  $\delta\phi$ of $g$-tensors in strained Ge. We use $L_z=20$~nm, $L_x=50$~nm, and squeeze QDs in $y$-direction. Squeezed QDs have generally lower relative variability.
    (c)-(d) Variability in unstrained Ge with $L_z=14$~nm, where tighter squeezing strongly pins the $g$-tensor and globally reduces variability, especially when $\textbf{B}$ is aligned to the long-QD direction. We exclude circular QDs ($L_x=L_y=50$~nm) in unstrained Ge because of their arbitrary angular fluctuations, caused by the random breaking of the QD symmetry. All results are obtained averaging over 200 disorder realizations.    }
    \label{fig:g AR vs phit}
\end{figure}

\paragraph{Circular QDs.--}
We begin by analyzing the variability of Ge hole qubits confined in circular QDs, where the in-plane confinement is on average isotropic with $\omega_x=\omega_y$ as expected from current device designs~\cite{hendrickxFourqubitGermaniumQuantum2021,johnTwodimensional10qubitArray2025a,Hendrickx2024}. 
Although the gate-induced confinement is isotropic in-plane, the symmetry of the QD is naturally broken by the electrostatic disorder and fluctuations of the confinement, that randomly deform the wavefunction according to the  local charge distribution and tuning of gate potentials.

In Fig.~\ref{fig:g vs Lz}, we compare the hole spins defined in strained Ge [Fig.~\ref{fig:g vs Lz}(a,b,c)] and unstrained Ge [Fig.~\ref{fig:g vs Lz}(d,e,f)] when the magnetic field is in-plane. Explicitly, here we consider $\textbf{B}$  aligned at $\pi/4$ from the $x$-direction.
We first study the dependence of the relative variability of the g-factor $\delta g$ on the QD size. Figs.~\ref{fig:g vs Lz}(a) and~(d) show the standard deviation of the relative  $\delta g$ in Eq.~\eqref{eq:dg} against the electrically tuneable average confinement length $\ell=L_x=L_y$ and the heterostructure width $L_z$ computed by averaging $\delta g$ over 200 different disorder realizations. We consider a constant electric field $F_z=1$~MV/m and $F_z=2$~MV/m for the strained and unstrained heterostructure, respectively and observe that the former are less variable than the latter. While in both cases the variability generally decreases for more smoothly confined QDs, strained Ge shows a clear monotonic reduction with increasing in-plane and decreasing out-of-plane confinement lengths ($\ell$ and $L_z$, respectively) whereas in unstrained Ge the variability becomes non-monotonic as the quantum-well width $L_z$ increases. In particular, in unstrained Ge when $L_z\gtrsim 15$~nm,  the variability plateaus for harmonic confinements $\ell\gtrsim 40$~nm.

The non-monotonic behavior of $\delta g$ is likely caused by the change in sign of the amplitude of $g_0\cdot \textbf{B}$, which approaches zero in this regime~\cite{rimbach-russGaplessSingleSpinQubit2025}, thus modifying the effect of disorder-induced corrections. The electrostatic configuration causing $g_0\cdot \textbf{B}\approx 0$ for in-plane field is qualitatively described by~\cite{michalLongitudinalTransverseElectric2021,rimbach-russGaplessSingleSpinQubit2025,abadillo-urielHoleSpinDrivingStrainInduced2023}
\begin{equation}
\label{perturbative correction}
    g_{xx,yy}\approx \mp 3q \pm \frac{6\hbar^2}{m_0 \Delta_\text{LH}} (\lambda \langle k_{x,y}^2\rangle\ - \lambda'  \langle k_{y,x}^2\rangle\ ) \ ,
\end{equation}
where $\langle{k_i^2}\rangle=1/2L_i^2$ and the parameters $\lambda$ and $\lambda'$ depend on the $z$-confinement, see the SM~\cite{supplemental_material} for more details. We neglect in these formulas corrections from finite inhomogeneous strain. In unstrained Ge QDs the heavy-light hole energy gap $\Delta_\text{LH}$ is smaller and the second term of Eq.~\eqref{perturbative correction} becomes comparable to  the bare Zeeman contribution $3q$ for realistic QD sizes, leading to a sign change in $g_0$ for tighter confinements. We further note that random fluctuations in the harmonic confinement strength, implemented as the second term of Eq.~\ref{Shift and WF oscillation}, are the strongest contribution to the variability of both strained and unstrained QDs. An extensive comparison of the different contributions is provided in the SM~\cite{supplemental_material}.

To gain further insight into the variability, in Fig.~\ref{fig:g vs Lz}(b) and (c) [Fig.~\ref{fig:g vs Lz}(e) and~(f)], we analyze the probability distribution function (PDF) of the g-factor variations $\delta g$ for QDs of different planar harmonic lengths $\ell$ confined in strained [unstrained] Ge wells of width $L_z=10$~nm and  $L_z=14$~nm, respectively.
The distribution becomes narrower for larger QDs, confirming that smoother confinement mitigates variability for both strained and unstrained Ge.

Interestingly, the statistical distribution of variability tends to be larger in unstrained Ge. Intuitively, the increased variability in unstrained and wider heterostructures can be understood from the larger SOI that develops in these cases. Larger SOI can improve the efficiency of the driving, but at the same time can also enhance the electrical susceptibility of $g$, increasing the variability of the qubit energy. This effect is particularly relevant in circular QDs where the electrostatic disorder breaks the symmetry in an uncontrolled way. 

\paragraph{Squeezed QDs.--}
By electrically squeezing the QD~\cite{boscoSqueezedHoleSpin2021}, we selectively imprint a preferred spin-quantization direction, enabling not only efficient and electrically tunable $g$-tensor engineering, but also a controlled reduction of variability for selected magnetic-field orientations.  
As shown in Fig.~\ref{fig:placeholder}(b)-(c), enforcing an elliptical confinement generally enhances the $g$-factor along the smooth confinement axis, while decreasing it along the transversal strongly-confined direction. This electrical control at the single-QD level enables the design of $g$-tensors arrays with arbitrary quantization-axis tilts, as required for hopping-based operations~\cite{Wang2024Hopping,lossQuantumComputationQuantum1998} and optimal multi-qubit gates~\cite{geyerAnisotropicExchangeInteraction2024,PhysRevB.109.085303,kp8s-py9m}.
Interestingly, this trend reverses in narrower Ge quantum wells when $L_z\lesssim L_{i}$, where larger $g$-factors are reached along the tight in-plane confinement direction, see SM~\cite{supplemental_material}.

Figs.~\ref{fig:placeholder}(b)-(c)  compare $g$-tensors in strained and unstrained Ge. 
In strained Ge, unpredictable long-range strain patterns arising from the substrate~\cite{corley-wiciakNanoscaleMapping3D2023} force the quantization axis of the $g$-tensor away from the preferential direction set by the electrostatic confinement. Here, we consider a strain field tilting the $g$-tensor by $\pi/4$. For circular QDs, unpredictable strain completely determines the anisotropy of the spin response, whereas in squeezed QDs electrostatic confinement partially stabilizes it. 
In contrast, unstrained Ge is lattice-matched and unaffected by long-range epitaxial strain fluctuations~\cite{costaBuriedUnstrainedGermanium2025}, experiencing only small inhomogeneous strain from the metal gates. As a result, the $g$-tensor is more reliably pinned to the direction set by the electrostatic confinement. 

Importantly, the pinning of $g$-tensors of squeezed QDs in unstrained Ge remains robust against charge disorder and inhomogeneous gate-induced strain, eliminating the need for extensive calibration across large-scale devices. 
Fig.~\ref{fig:g AR vs phit} examines the variability of the spin response in QDs elongated along the $x$-direction, focusing on fluctuations in both Zeeman energy amplitude and tilt angle of $g$.
In Figs.~\ref{fig:g AR vs phit}(a)-(c), we compare the relative change of the Zeeman energy in strained and unstrained Ge QDs with different aspect ratios against in-plane $\textbf{B}$ field angle $\phi_B$. In both types of devices, the variability is minimized when $\textbf{B}$ is aligned with the long confinement direction and generally increases when it approaches the strong-confinement axis. This behavior is most reliable in unstrained Ge, where $\delta g$ becomes negligibly small in strongly squeezed QDs at $\phi_B= 0, \pi$. In contrast, strained Ge exhibits strong modulations induced by long-range strain.

Squeezed QDs are also robust against tilts of $g$-tensors, enabling precise and reliable $g$-tensor engineering. Figs.~\ref{fig:g AR vs phit}(b)-(d) examine the angular variability of $g$, by showing the PDF of the angle rotation $\delta\phi$ between $g$-tensors with and without disorder. Both strained and unstrained Ge exhibit sharply peaked distributions when the confinement is strongly anisotropic, pinning $g_N$ within $1^o$ to  $5^o$ of $g_0$. While unstrained Ge devices respond monotonically to the increase of aspect ratio of the electrostatic confinement, strained devices behave poorly for intermediate confinement strengths and only outperform the isotropic case for large aspect ratios. This can be intuitively understood by noting that in strained devices an elliptical electrostatic confinement performs a rotation on the g-tensor, moving it from the predefined direction of the strain field to the electrostatically imprinted one. Consequently, weakly elliptical QDs with fluctuations in the harmonic confinement strengths trigger unwanted rotation between the circular and the highly elliptical one. These results highlight that squeezed QDs, especially in unstrained Ge, constitute a promising route toward reproducible and robust qubit properties in large-scale devices.

\paragraph{Conclusion.--}
In conclusion, we show that electrically squeezing QDs not only enables a robust way to on-demand engineer $g$-tensor, but also provides a controllable way to reduce the variability of the spin response in both strained and unstrained Ge QDs at selected $\textbf{B}$-field directions. In particular, we find minimal magnitude and angular variability of the $g$-tensor in smoothly-confined QDs in unstrained Ge when $\textbf{B}$ is aligned to the long in-plane confinement axis, providing design guidelines for scalable and reproducible Ge-based spin qubit processors.

\paragraph{Acknowledgments.-- }
We thank all members of the Bosco, Rimbach-Russ, Scappucci, Veldhorst, and Vandersypen group for valuable feedback. This research was supported by the H2024 QLSI2 project, the Army Research Office under Award Number: W911NF-23-1-0110, and NCCR Spin (grant number 225153). 
The views and conclusions contained in this document are those of the authors and should not be interpreted as representing the official policies, either expressed or implied, of the Army Research Office or the U.S. Government. The U.S. Government is authorized to reproduce and distribute reprints for Government purposes notwithstanding any copyright notation herein.

\bibliography{references}

\clearpage
\newpage
\mbox{~}

\onecolumngrid

\setcounter{equation}{0}
\setcounter{figure}{0}
\setcounter{table}{0}
\setcounter{section}{0}

\renewcommand{\theequation}{S\arabic{equation}}
\renewcommand{\thefigure}{S\arabic{figure}}
\renewcommand{\thesection}{S\arabic{section}}
\renewcommand{\bibnumfmt}[1]{[S#1]}

\begin{center}
  \textbf{\large Supplemental Material:\\
  Electrically Tuneable Variability in Germanium Hole Spin Qubits}\\[.2cm]
 Edmondo Valvo$^{1,*}$, Michèle Jakob$^{1,*}$, Patrick Del Vecchio$^{1}$, Maximilian Rimbach-Russ$^{1}$ Stefano Bosco$^{1}$ \\[.1cm]
   {\itshape $^1$QuTech and Kavli Institute of Nanoscience, Delft University of Technology, Delft, The Netherlands} 
\end{center}
\maketitle

\section*{Abstract}
In this Supplemental Material, we  provide explicit expressions of the Luttinger-Kohn and  Bir-Pikus Hamiltonian and describe the inhomogeneous strain fields  used in the main text. We also give additional details on the pristine g-tensor, including the inversion of the easy and hard quantization axes in squeezed dots for different heterostructure widths. Finally, we compare the various sources of variability and show that, while a typical number of positive or negative charge traps can substantially affect the angular tilt of the g-tensor, amplitude variations are dominated by changes in the dot size.

\section{Theoretical model} 
\label{HamiltonianSupplementary}
Hole qubits in semiconductors and gate-defined quantum dots are well described by the Luttinger-Kohn-Bir-Pikus
Hamiltonian. In this section, we provide the full expression of the Luttinger-Kohn, $H_{LK}$ and Bir-Pikus, $H_{BP}$ Hamiltonians.
We also discuss in more detail the inhomogeneous strain field used in simulations of the main text.
\subsection{Luttinger-Kohn and Bir Pikus}
The Luttinger-Kohn Hamiltnonian describes the spin-$3/2$ and spin-$1/2$ holes near the $\Gamma$-point and is given by
\begin{equation}
\label{LK4by4}
\begin{aligned}
     H_{LK}= \frac{\hbar^2}{2 m_0}& \Big[(\gamma_1 +\frac{5 \gamma_2}{2}) \frac{\textbf{k}^2}{2} - \gamma_2  \sum_{i}(k_i^2 J_i^2) - 2 \gamma_3 \sum_{i<j} \{k_i,k_j\}\{J_i,J_j\}\Big],
\end{aligned}
\end{equation}
where the dimensionless parameters for Germanium are  $\gamma_1 = 13.35$, $\gamma_2= 4.24$ and $\gamma_3 = 5.69$ \cite{Terrazos2021}. 
The coordinates $x, \ y, \ z$ are aligned to the crystallographic axes [100], [010], [001], respectively.
 
The impact of strain is described by the Bir-Pikus Hamiltonian, which contains in its general formulation both diagonal and off-diagonal components 
\begin{equation}
        H_{BP} = -\left(a+\frac{5b}{4} \right)\ \text{Tr}(\epsilon) + b \sum_i \epsilon_{ii} J_i^2  +\frac{2d}{\sqrt{3}} \left( e_{xy} \{ J_x, J_y \} + c.p. \right),
\end{equation}
where $a, \ b$, and $d$ are material dependent deformation potentials. 
Explicitly, in the basis $(\text{HH}\uparrow, \ \text{HH} \downarrow, \ \text{LH}\uparrow, \ \text{LH}\downarrow)$ the Hamiltonian is given by
\begin{equation}
\label{eq:HBP}
    \begin{aligned}
   H_{BP}  &= \begin{pmatrix}
         -a \epsilon_A -\frac{1}{2}b \epsilon_D && 0 && d (\epsilon_{xz}- i \epsilon_{yz}) && -id \epsilon_xy + \sqrt{3} b \ \epsilon_p \\
         0 && -a \epsilon_A-\frac{1}{2}b \epsilon_D && i d \epsilon_{xy} + \sqrt{3}b\ \epsilon_p && -d(\epsilon_{xz} + i \epsilon_{yz})\\
         d(\epsilon_{xxz}+i\epsilon_{yz}) && -i d\epsilon_{xy} + \sqrt{3}b\ \epsilon_p && -a \epsilon_A+\frac{1}{2}b \epsilon_D && 0 \\
         id \epsilon_{xy} + \sqrt{3}b\ \epsilon_p && -d (\epsilon_{xz}- i \epsilon_{yz}) && 0 && -a \epsilon_A+\frac{1}{2}b \epsilon_D\\
     \end{pmatrix}\\
     &= -a \epsilon_A\sigma_0 \sigma_0+\epsilon_D\ \frac{1}{2} b \sigma_x \sigma_0 + \epsilon_p\ \sqrt{3} b \sigma_x \sigma_x +\epsilon_{xy}\ id \sigma_y \sigma_y +\epsilon_{xz}\ d \sigma_x \sigma_z +\epsilon_{yz}\ d \sigma_y \sigma_0\\
     &=-a \epsilon_A I+ H_{\epsilon_D}+H_{\epsilon_p} + H_{\epsilon_{xy}}+H_{\epsilon_{xz}}+ H_{\epsilon_{yz}} \ ,
\end{aligned}
\end{equation}
where we introduce $\epsilon_A = \epsilon_{xx}+\epsilon_{yy}+\epsilon_{zz}$, $\epsilon_D=\epsilon_{xx}+ \epsilon_{yy}-2\epsilon_{zz}$ and $\epsilon_p=\epsilon_{xx}- \epsilon_{yy}$. Because the the strain spatial variation underneath the plunger gate are approximately constant~\cite{abadillo-urielHoleSpinDrivingStrainInduced2023}, the $-a \epsilon_A$ component leads to a global energy shift and we neglect it in the main text.

\subsection{Inhomogeneous strain fields}
\label{sec:spatialStrain}
\begin{figure}[h!]
\centering
\includegraphics[width=0.9\textwidth]{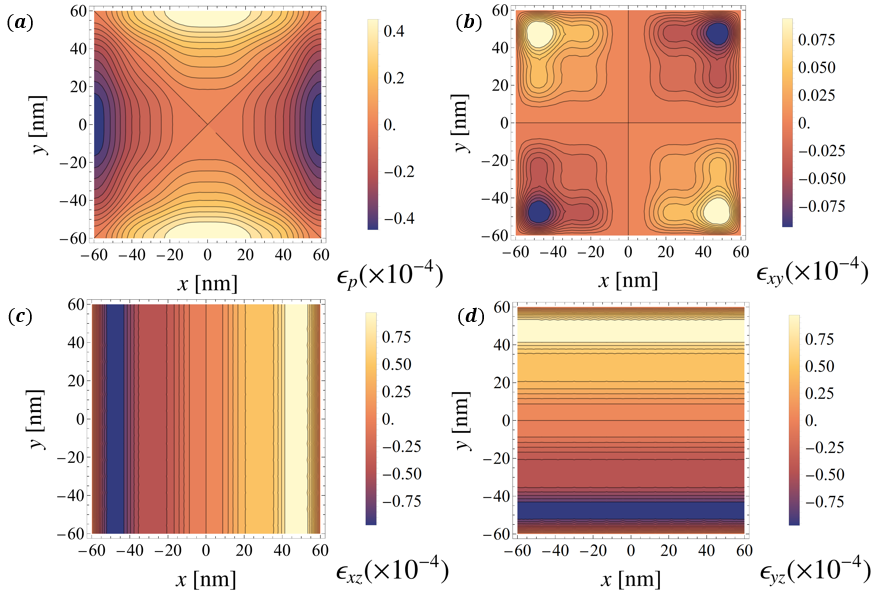}
\caption{Here we show how the strain varies in the $xy$-plane underneath the plunger gate, with radius ${L}_p = 60\ \rm nm$, for the different strain components used in our simulations.}
\label{fig:first}
\end{figure}
Following Ref.~\cite{abadillo-urielHoleSpinDrivingStrainInduced2023}, we model the in-plane spatial dependencies of the different elements of the Bir-Pikus Hamiltonian with $\epsilon_p = -(x^2-y^2)$, $\epsilon_{xz}= x$, $\epsilon_{yz}= y$ and $\epsilon_{xy}= -xy$. To bound the strain values in-plane, we perform a Fourier transform with respect to the radius of the top plunger gate and consider only Fourier components up to 4th order. Since all of these functions are separable, we can write the explicit function of the spatial dependency of the strain as
\begin{equation}
    \begin{split}
        \epsilon_p (x,y) =& \varepsilon_p\Biggl(-144 \cos \left(\frac{\pi  x}{L_p}\right)+36 \cos \left(\frac{2 \pi  x}{L_p}\right)-16 \cos \left(\frac{3 \pi  x}{L_p}\right)+9 \cos \left(\frac{4 \pi  x}{L_p}\right)\\
        &+144 \cos \left(\frac{\pi  y}{L_p}\right)-36 \cos \left(\frac{2 \pi  y}{L_p}\right)+16 \cos \left(\frac{3 \pi  y}{L_p}\right)-9 \cos \left(\frac{4 \pi  y}{L_p}\right)\Biggr) \ ,\\
        \epsilon_{xy} (x,y)=& \varepsilon_{xy} \Bigg[\left(12 \sin \left(\frac{\pi  x}{{L}_p}\right)-6 \sin \left(\frac{2 \pi  x}{{L}_p}\right)+4 \sin \left(\frac{3 \pi  x}{{L}_p}\right)-3 \sin \left(\frac{4 \pi  x}{{L}_p}\right)\right)\\
        &\times \left(-12 \sin \left(\frac{\pi  y}{{L}_p}\right)+6 \sin \left(\frac{2 \pi  y}{{L}_p}\right)-4 \sin \left(\frac{3 \pi  y}{{L}_p}\right)+3 \sin \left(\frac{4 \pi  y}{{L}_p}\right)\right)\Bigg] \ ,\\
        \epsilon_{xz}(x,z)=&\varepsilon_{xz} (1+z) \left(12 \sin \left(\frac{\pi  x}{{L}_p}\right)-6 \sin \left(\frac{2 \pi  x}{{L}_p}\right)+4 \sin \left(\frac{3 \pi  x}{{L}_p}\right)-3 \sin \left(\frac{4 \pi  x}{{L}_p}\right)\right) \ ,\\
        \epsilon_{yz}(y,z)=& \varepsilon_{yz}\ (1+ z) \left(12 \sin \left(\frac{\pi  y}{{L}_p}\right)-6 \sin \left(\frac{2 \pi  y}{{L}_p}\right)+4 \sin \left(\frac{3 \pi  y}{{L}_p}\right)-3 \sin \left(\frac{4 \pi  y}{{L}_p}\right)\right).
    \end{split}
\end{equation}
In this work, we use $\varepsilon_p = - \left(\frac{1}{6 \pi}\right)^2 \times \frac{1}{2} \cdot 10^{-4}$, $\varepsilon_{xy}=-\left(\frac{1}{6 \pi}\right)^2 \times 10^{-5}$, $\varepsilon_{xz}=\left(\frac{1}{6 \pi}\right) \times 10^{-4}$ and $\varepsilon_{yz}=\left(\frac{1}{6 \pi}\right) \times 10^{-4}$, qualitatively matching Ref.~\cite{abadillo-urielHoleSpinDrivingStrainInduced2023}. We consider a plunge gate with radius $L_p = 60\ \rm nm$.
These inhomogeneous strain fields are shown in Fig.~\ref{fig:first}.

The long-range strain fluctuations induced by the substrate are included as a constant term, over the size of the QD. They are modelled with the Bir-Pikus Hamiltonian  given in Eq.~(\ref{eq:HBP}) with $\varepsilon_{xx} = -0.0061= \varepsilon_{yy}$, $\varepsilon_{zz}=0.001$, $\varepsilon_{xy}=10^{-5}$ and $\varepsilon_{xz}=\varepsilon_{yz}=10^{-4}$.

\section{Engineering the g-tensor of squeezed dots}
\label{perturbation g-tensor}
\begin{figure}[h!]
    \centering
    \includegraphics[width=0.8\linewidth]{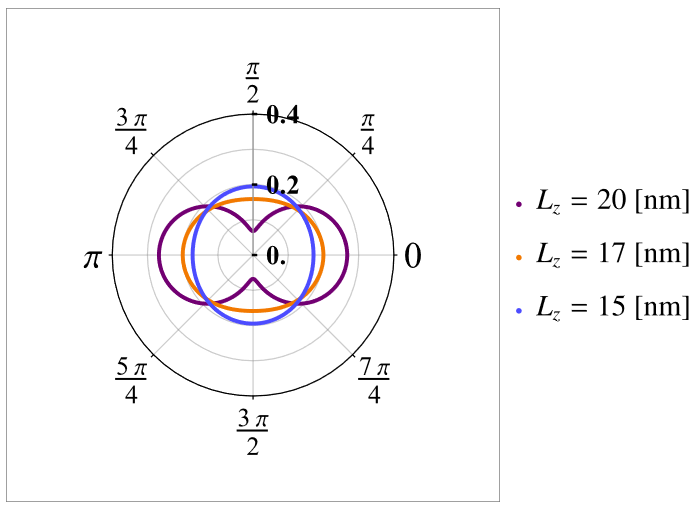}
    \caption{Inversion point in unstrained germanium. The figure presents the $g$-tensor of a quantum dot with lateral confinement lengths $L_x = 50\ \mathrm{nm}$ and $L_y = 20\ \mathrm{nm}$, evaluated for varying germanium well thicknesses $L_z$. For sufficiently large well thicknesses, the $g$-tensor is pinned along the $x$ direction. As the germanium well thickness is reduced, this pinning is inverted, and the principal axis of the $g$-tensor becomes aligned with the more strongly confined (squeezed) lateral direction of the quantum dot}
    \label{fig:zero-point}
\end{figure}

In this section, we analyze the peculiar behavior of the g-tensor of squeezed dots as a function of well thickness in the unstrained germanium device. Fig.~\ref{fig:zero-point} shows the g-tensor for a squeezed quantum dot with lateral confinement lengths $L_x=50$~nm and  $L_y=20$~nm with varying germanium well thickness $L_z\in [20,17,15,13]$~nm. 
We can clearly see that between $L_z=15$~nm and $17$~nm the g-tensor changes its response with respect to an elliptical squeezing. For thinner wells, the g-tensor inverts its response to the electrostatic confinement compared to the main text and becomes larger along the direction of strongest confinement. From the analytical expression of the g-tensor

\begin{equation}
    g_{xx,yy} \approx \mp  3q \pm \frac{6\hbar^2}{m_0 \Delta_\text{LH}} (\lambda \langle k_{x,y}^2\rangle\ - \lambda'  \langle k_{y,x}^2\rangle\ ) \ .
\end{equation}

Assuming that the groundstate wavefunction is separable in $z$ and $x,y$ direction, explicit expressions of the constants above are~\cite{michalLongitudinalTransverseElectric2021,abadillo-urielHoleSpinDrivingStrainInduced2023} 
\begin{align}
    \lambda &= 2\eta_h \gamma_3^2-\widetilde{\kappa} \gamma_2\\
    \lambda^\prime &= 2\eta_h \gamma_3 \gamma_2-\widetilde{\kappa} \gamma_2\\
    \widetilde{\kappa} &=\kappa - 2\gamma_3\widetilde{\eta}_h\\
    \gamma_h &= \frac{6\gamma_3^2\hbar^2}{m_0}\sum_{l} \frac{\left|\langle{\text{HH}_0}| k_z | {\text{LH}_l}\rangle \right|^2}{E_{\text{LH},l}-E_{\text{HH},0}}\\
    \eta_h &= -\Delta_\text{LH}\sum_{l} \frac{2\text{Im}\left(\langle{\text{HH}_0}|z |{\text{LH}_l}\rangle\langle{\text{LH}_l}|k_z|{\text{HH}_0}\right)\rangle}{E_{\text{LH},l}-E_{\text{HH},0}}\\
    \widetilde{\eta}_h &= \Delta_\text{LH}\sum_{l} \frac{\text{Im}\left(\langle{\text{HH}_0}|k_zz+zk_z|{\text{LH}_l}\rangle\langle{\text{LH}_l|\text{HH}_0}\rangle\right)}{E_{\text{LH},l}-E_{\text{HH},0}}.
\end{align}
 we  see that once the spin-orbit correction term becomes larger than $3q$ any additional increase in $\langle k_{y,x}^2\rangle$ only enlarges the absolute value of the g-factor instead of shrinking it.
\section{Pristine g-tensor and variability contributions}
In this section, we provide additional information on the pristine $g$-tensor $g_0$ and on the role of different disorder contribution on the variability.
\subsection{Pristine g-tensor of circular dots}
\label{pristinegtens}
\begin{figure}[h!]
    \centering
    \includegraphics[width=0.8\linewidth]{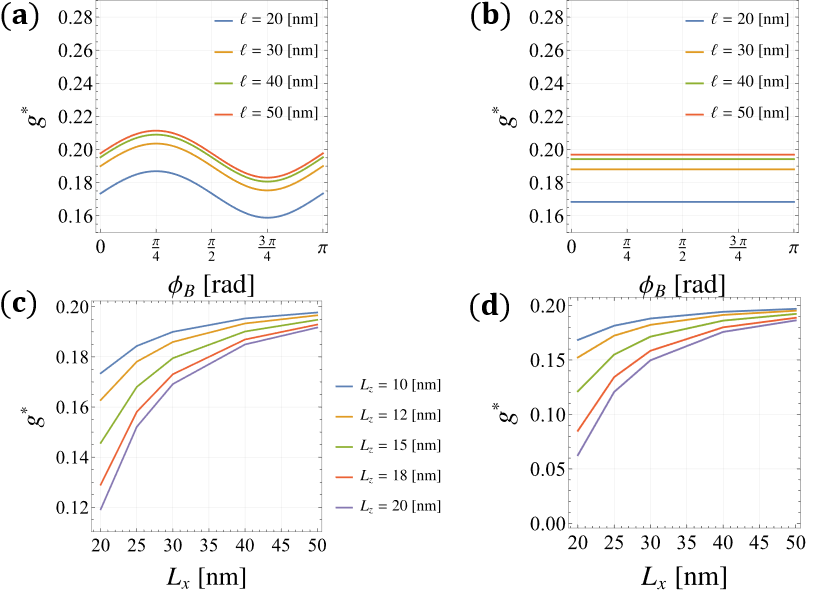}
    \caption{Simulations of the pristine g-factor of circular QDs for (a) strained and (b) unstrained germanium wells with Ge well width and confinement length $L_z=10$~nm well for varying planar confinements as a function of magnetic field angle. Planar g-factor at $\phi_B=0$ for (c) strained and (d) unstrained wells respectively as a function of the  confinement length $L_x$ and for different Ge well thicknesses $L_z$. Unless otherwise mentioned, we use the same simulation parameters and model as in the main text.}
    \label{fig:planar_g0}
\end{figure}
 We report here the pristine $g$-tensor $g_0$ in the cases examined in the main text. We define here the g-factor $g^*$, related to the pristine $g$-tensor $g_0$ by $g^*=g_0\textbf{B}/|\textbf{B}|$. As seen in Fig.~\ref{fig:planar_g0}, the planar g-factor of circular dots in strained germanium roughly varies between $0.12$ and $0.2$ an anisotropy imprinted by the long-range strain fluctuations~\cite{corley-wiciakNanoscaleMapping3D2023}. In contrast, without disorder holes in unstrained germanium are symmetric in-plane and $g^*$ remains constant as $\textbf{B}$ rotates by an angle $\phi_B$. By considering a well thickness of $L_z=10$~nm $g^*$  of strained and unstrained Ge fall into similar ranges. In Figs.~\ref{fig:planar_g0}(c) and (d), we show $g^*$ against the size of the dot $L_x$ for strained and unstrained Ge when $\textbf{B}$ is aligned to the x-direction ($\phi_B=0$) . In smaller dots, we observe a strong suppression of the planar g-factor, especially in unstrained heterostructure, caused by the  
 
\begin{figure}[h!]
    \centering
    \includegraphics[width=0.8\linewidth]{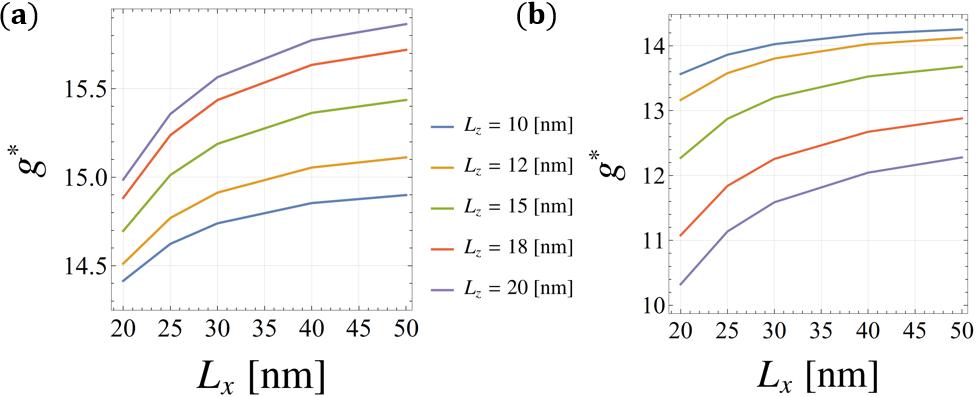}
    \caption{Out of plane g-factor of circular dots in strained (a) and unstrained (b) Ge wells. We show $g^*$ when $\textbf{B}$ is perpendicular to the substrate as a function of the in-plane confinement $L_x=L_y$ and for different values of heterostructure widths $L_z$.}
    \label{fig:perp_g0}
\end{figure}

 For completeness, in Fig.~\ref{fig:perp_g0}, we report the out-of-plane g-factor. We observe a small reduction of $g^*$  for unstrained Ge comapared to strained Ge, and interestingly the opposite trend as a function of well thicknesses, decreasing (increasing) in unstrained (strained) Ge for thicker wells. With these pristine g-tensor values and assuming a magnetic field of $10$~mT, the expected variations in the Zeeman energy of the considered spin qubit fall between $0.1$~MHz at points of low variability up to $10$~MHz in regions of particularly large variability.

\subsection{Noise contributions comparison}
\begin{figure}[h!]
    \centering
    \includegraphics[width=\linewidth]{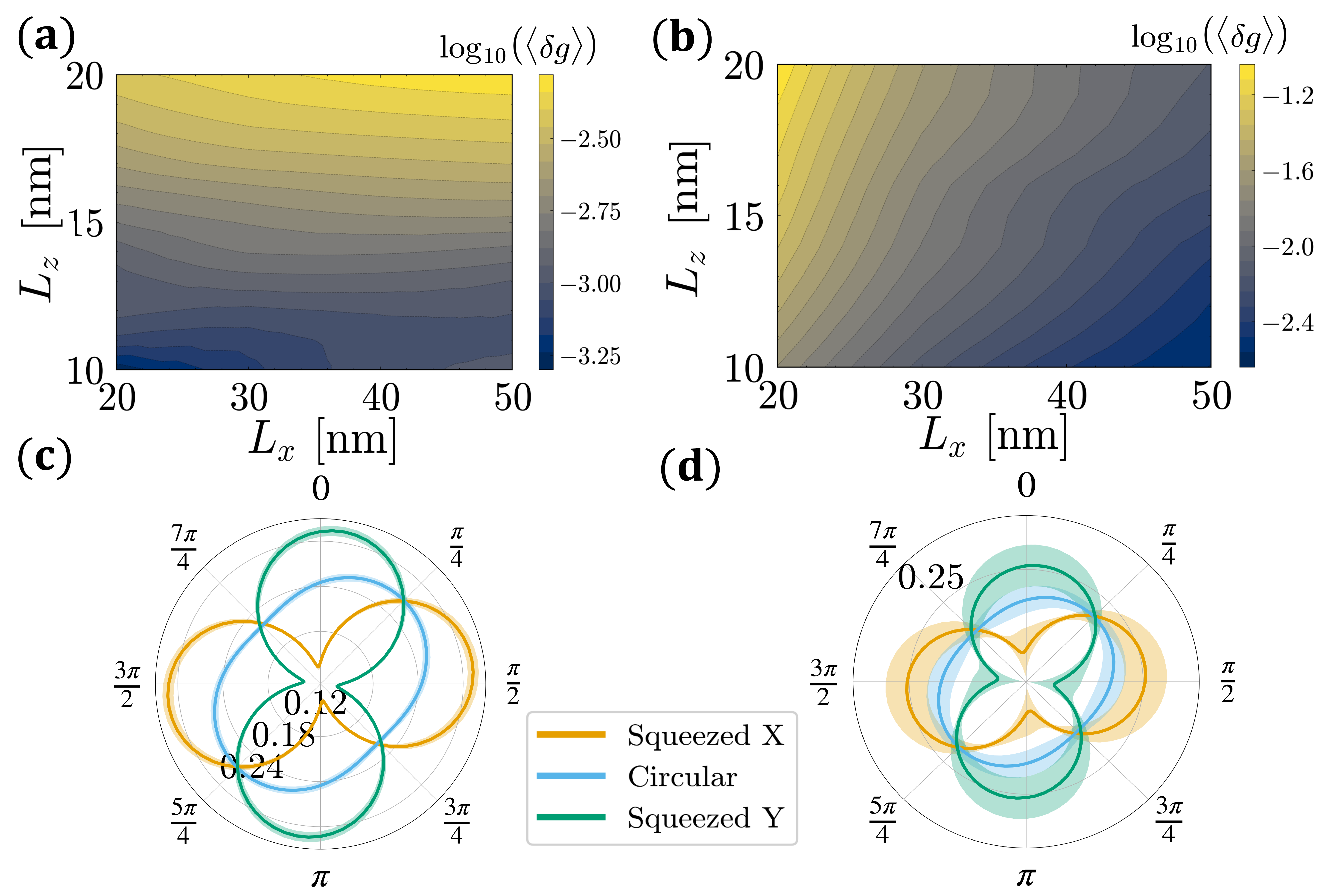}
    \caption{Standard deviation of the distribution of g-factor variation for strained  wells with only interface disorder (a) and with added harmonic confinement fluctuations (b). Mean of the disordered g-tensors with the standard deviation on top including only interface disorder (c) and the full disorder configuration (d). We use the same simulation parameters as in the main text.}
    \label{fig:NoiseComparison}
\end{figure}

In this section, we compare the impact of interface disorder and fluctuations in the harmonic confinement. In particular, we find that considering realistic values of interface disorder~\cite{paqueletwuetzReducingChargeNoise2023}, the impact of interface disorder on the g-factor amplitude is almost one order of magnitude smaller than the variability induced by fluctuations in the harmonic confinement. In Fig.~\ref{fig:NoiseComparison}~(a)-(c), we show an analogous plot as in the main text considering only interface disorder, while in  Fig.~\ref{fig:NoiseComparison}~(b)-(d) we add  random fluctuations of the planar confinement. By comparing Fig.~\ref{fig:NoiseComparison} (a) and (b), we observe that the addition of harmonic confinement fluctuations confirms the detrimental effect of large Ge wells. However, in this case we find the opposite trend with respect to planar confinement strengths. 

In particular, including electrostatic disorder wider dots are more resilient to variability than tightly confined dots. This can be qualitatively understood by recalling the perturbative corrections to the planar g-tensor of circular dots ($\ell=L_x=L_y$)~\cite{abadillo-urielHoleSpinDrivingStrainInduced2023,rimbach-russGaplessSingleSpinQubit2025}
\begin{equation}
    g^* \approx 3q - \frac{6\hbar^2}{m_0 \Delta_\text{LH}} 2 \eta_h (\gamma_3 ^2 - \gamma_3\gamma_2 ) \langle k^2\rangle ,
\end{equation}
with $\eta_h$ being a parameter dependent on confinement along the $z$-direction and $ \langle k^2\rangle=1/2L_x^2$. If we vary $L\to L(1+\delta)$ by considering small fluctuations of the confinement, then then changes in  $g^*$ are  $\propto \delta /\Delta_\text{LH}L^2$, thus explaining the lower effect of electrostatic variability for large dots (larger $L_x$) and for strained Ge (where $\Delta_\text{LH}$ is larger).

\begin{figure}[h!]
    \centering
    \includegraphics[width=\linewidth]{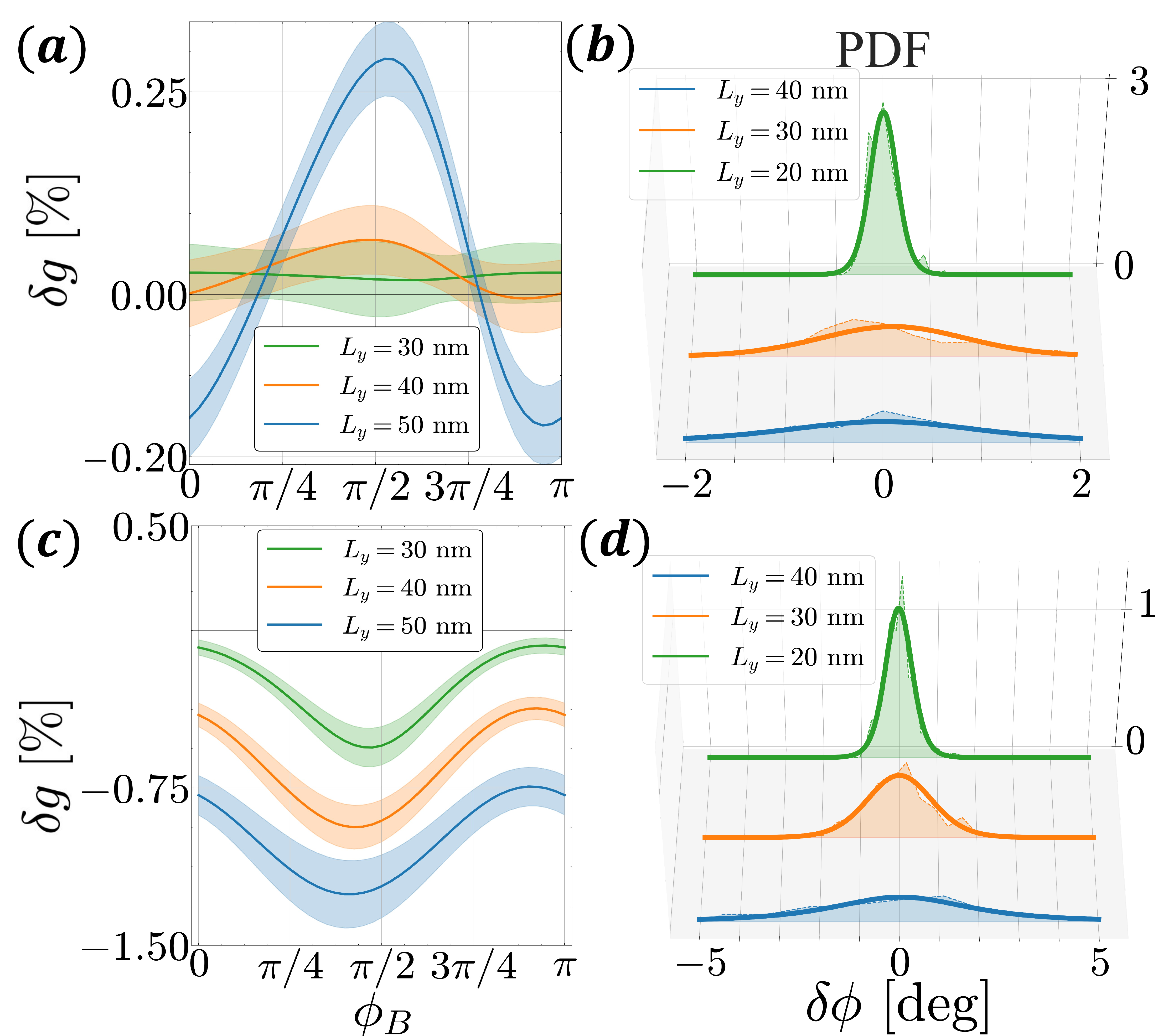}
    \caption{ (a) Simulated variability of the relative $g$-factor $\delta g$ for different $\textbf{B}$ field directions $\phi_B$ and (b) probability density function of the tilt angle  $\delta\phi$ of $g$-tensors in strained Ge. We use $L_z=20$~nm, $L_x=50$~nm, and squeeze QDs in $y$-direction. Squeezed QDs have generally lower relative variability.
    (c)-(d) Variability in unstrained Ge with $L_z=14$~nm, where tighter squeezing strongly pins the $g$-tensor and globally reduces variability, especially when $\textbf{B}$ is aligned to the long-confinement direction. All these plots are computed with only interface disorder and no harmonic confinement fluctuations.}
    \label{fig:OldFigure3}
\end{figure}

Furthermore, we study the effects of interface disorder on squeezed dots in Fig.~\ref{fig:OldFigure3}. In (a) and (c) we analyze the effects of only charge disorder against the full electrostatic disorder (b) and (d) [see Fig.3 in the main text]. We  observe a steep decrease in amplitude variability if only charge disorder is included. There is also a decrease in angular variability that is not as large, but still reduces $\delta \phi$ within $2^\circ-5^\circ$ of the pristine values. Moreover, without fluctuations of harmonic confinement, the strained Ge dot presents a monotonic dependence of the distribution spread with respect to the confinement strength. Fundamentally, this comparison highlights that imperfect control of the harmonic confinement is a critical problem and for relatively clean interfaces, it overcomes random interface disorder for static g-tensor engineering.
\newpage
\subsection{Disorder from positive interface charges}
To confirm the general applicability of our analysis for device working in both depletion and accumulation mode, we have repeated our simulations for a few representative cases using positive charges instead of negative charges from the main text. Charge disorder with positive charges can arise in systems if more holes are trapped close to the interface as negative charges in the oxide~\cite{sangwanImpactSurfaceTreatments2025}. 
\begin{figure}[h!]
    \centering
    \includegraphics[width=\linewidth]{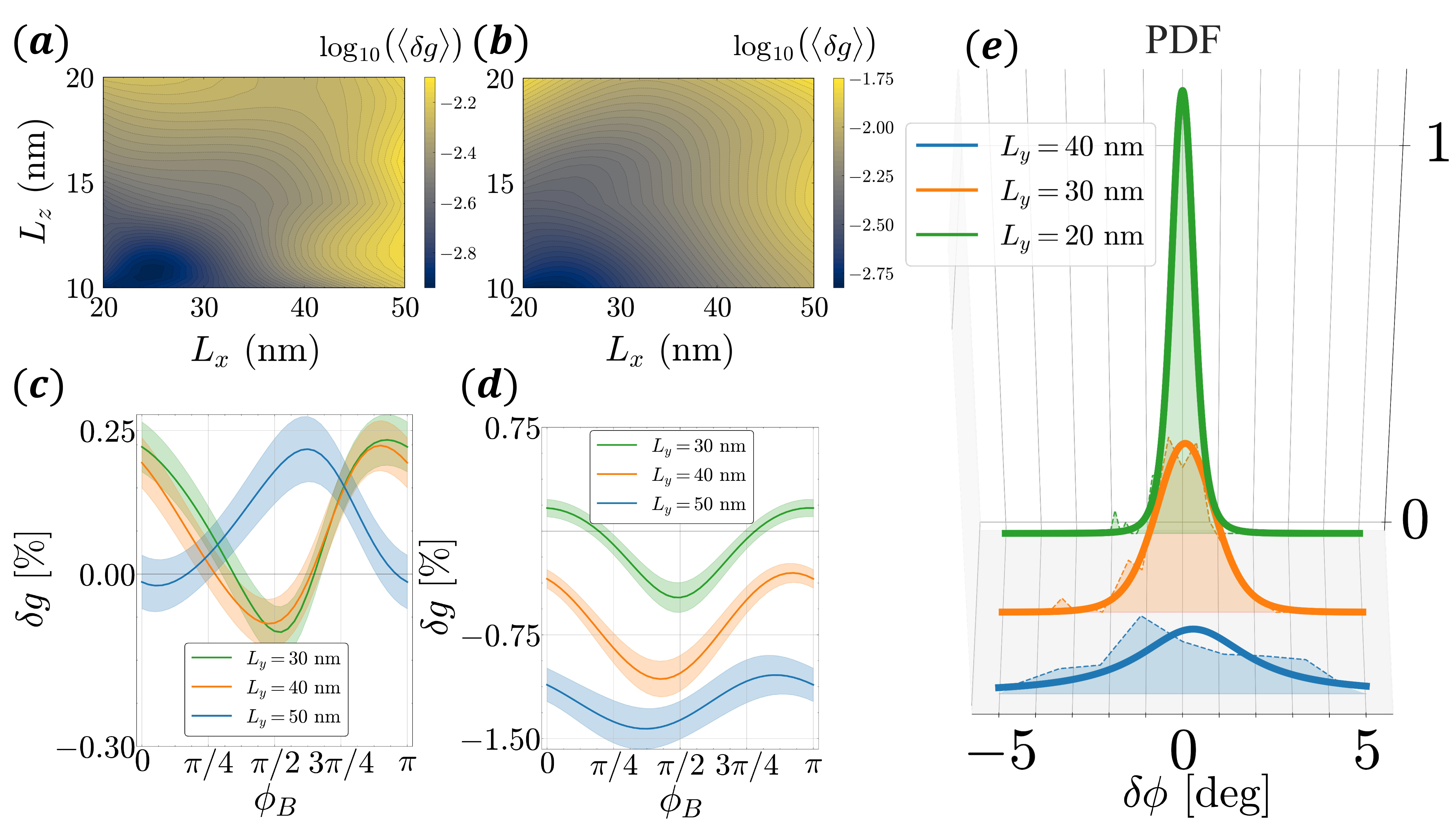}
    \caption{ Simulated g-factor variation for (a) strained and (b) unstrained germanium wells for positive interface charges for circular QDs.  Variability of squeezed dots for a strained (c) and unstrained (d) germanium well of width $L_z=20$~nm. (e) Distribution of angle tilt for an unstrained germanium well device with $L_z=14$~nm.}
    \label{fig:PositiveCharge}
\end{figure}
Fig.~\ref{fig:PositiveCharge} shows the g-tensor variability as reported in the main text (see also Fig.~\ref{fig:OldFigure3}) for positively charged defects. We observe an almost indistinguishable behavior; more tightly confined dots present the same narrow distributions for both magnitude and angular variability, highlighting the general applicability of our analysis to both depletion and accumulation Ge devices.

\end{document}